\documentstyle[aps,prl,epsf,floats,twocolumn]{revtex}
\begin{document}
\draft
\twocolumn[\hsize\textwidth\columnwidth\hsize\csname @twocolumnfalse\endcsname
]
\noindent {\bf Comment on ``Electric Field Scaling at $B=0$ Metal-Insulator
Transition in Two Dimensions''}
\vspace{1pt}

In a recent Letter, Kravchenko et al.~\cite{Krav} have provided evidence for
a metal-insulator transition (MIT) in a two-dimensional electron system
(2DES) in Si metal-oxide-semiconductor field-effect transistors (MOSFETs).  The
transition observed in these samples~\cite{Krav,mit1} occurs at relatively
low electron densities $n_{s}\sim (1-2)\times 10^{11}$cm$^{-2}$ so that the
electron-electron interaction $U\propto\sqrt n_{s}$ is about an order of 
magnitude greater than the Fermi energy $E_{F}\propto n_s$, suggesting that 
this MIT is driven by electron-electron interactions.  
However, the disorder in
Si MOSFETs at the transition is also strong, as indicated by the 
reported~\cite{Krav,mit1} values of the critical conductivity 
$\sigma_{c}\sim e^{2}/2h$.  

In this Comment, we present evidence for a 2D MIT in another Si-based
structure where the disorder, as measured by the mobility, is about two orders
of magnitude weaker than in Si MOSFETs.  We find that the MIT occurs in the
same range of $n_s$ as in Si MOSFETs.  This provides clear and strong
evidence that the 2D MIT in Si-based devices is caused by electron-electron 
interactions.  

The measurements were carried out on a 2DES located in a 10~nm thick Si channel
sandwiched between a Si$_{0.75}$Ge$_{0.25}$ buffer ($\sim 1.5~\mu$m thick) and
a 15~nm thick Si$_{0.75}$Ge$_{0.25}$ spacer, which was followed by a 10~nm
thick Si$_{0.75}$Ge$_{0.25}$ supply layer ($N_{D}=4\times 
10^{18}$cm$^{-3}$), and a 4~nm thick Si cap layer ($N_{D}=4-10\times 
10^{18}$cm$^{-3}$)~\cite{APL}.  The samples were standard Hall bars with a 
source-to-drain length $L=100~\mu$m and a width $W=10~\mu$m.  Resistance $R$
was measured using ac lock-in techniques at temperatures $0.35<T<4.2$~K as a 
function of the carrier density $n_s$, which was varied using the front gate 
at a fixed back-gate bias $V_{BG}$.  The effect of the $V_{BG}$ was to change 
the mobility of a 2DES at a given $n_s$.

Fig.~1(a) shows the conductivity $\sigma =L/(WR)$ as a function of $n_s$ at
\begin{figure}[t]
\epsfxsize=2.9in \epsfbox{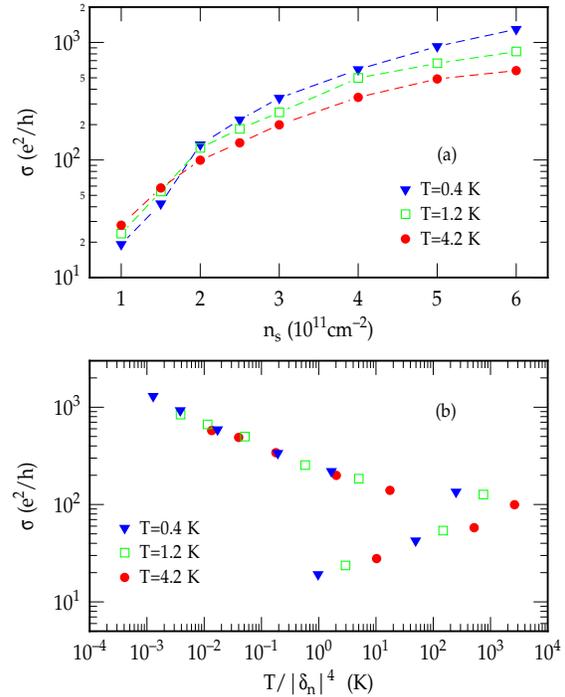}\vspace{5pt}
\caption{(a) $\sigma$ vs. $n_s$ for $T=4.2, 1.2, 
0.4$~K and $V_{BG}=10$~V.  The critical density $n_{c}\approx 1.8\times 
10^{11}$cm$^{-2}$.  (b) Scaling of $\sigma$ with $T$ using the data shown in 
(a).  $\delta_{n}\equiv (n_{s}-n_{c})/n_{c}$.}
\vspace{-12pt}
\end{figure}
several $T$.  Clearly, for $n_{s}>n_{c}\approx 1.8\times 
10^{11}$cm$^{-2}$, $\sigma$ exhibits the metallic $T$-dependence: it increases
as $T$ is lowered.  For $n_{s}<n_{c}$, the $T$-dependence is reversed, i.~e.
it is insulating.  Scaling of $\sigma$ with $T$~\cite{Krav,mit1} is shown in 
Fig.~1(b) with the scaling exponent of $4\pm 1$ but more
work is in progress in order to determine its value more precisely.
Even though these data look very similar to the results 
obtained on Si MOSFETs~\cite{Krav,mit1}, there is at least one
striking difference: $\sigma_{c}\sim 100~e^{2}/h$.
Therefore, it is clear that the value of $\sigma_c$ at a 2D 
MIT is {\em not} universal.  Very high mobilities have 
been reported earlier~\cite{APL} in these modulation-doped Si/SiGe 
heterostructures.  For example, the mobility $\mu =\sigma /n_{s}e$ at the
MIT in Fig.~1(a) is of the order of 100,000~cm$^{2}$/Vs, whereas 
$\mu\sim$1,000~cm$^{2}$/Vs at the MIT in Si MOSFETs~\cite{Krav,mit1}.  
Our most important result, however, is that the MIT in Si/SiGe
heterostructures occurs for $n_c$ comparable to those in
Si MOSFETs, so that $U$ at the MIT
in both types of samples is about the same.  On the other hand, our Si/SiGe
samples are very weakly disordered: $k_{F}l\approx 40$ ($k_F$ -- Fermi 
wavenumber, $l$ -- mean free path) at $n_{s}=n_{c}$ in these samples, as 
opposed to $k_{F}l${\scriptsize $\stackrel{\textstyle _<}{_\sim}$}1 in Si 
MOSFETs.  These results provide clear and convincing evidence that $U$ is the 
relevant energy scale at the MIT in Si-based devices, and not disorder.
Since Si is a multivalley semiconductor, it would be interesting to study 
this transition in other multivalley and single valley semiconductors.

The support by the NSF Grant No. DMR-9510355 (DP, SW), and of the ARO Grant 
No. 33036--EL (SW) is acknowledged.

\begin{center}
K. Ismail$^1$, J. O. Chu$^1$, Dragana Popovi\'{c}$^{2}$, \\ 
A. B. Fowler$^{1}$, and S. Washburn$^{3}$\\
\end{center}
\vspace{-6pt}
{\footnotesize
$^{1}$ IBM Research Division, T. J. Watson Research Center, Yorktown 
Heights, NY 10598 \\ $^{2}$ Department of Physics, City College of the City 
University of New York, New York, NY 10031, and National High Magnetic Field
Laboratory, Florida State University, Tallahassee, FL 32306 \\ $^{3}$Dept. of 
Physics and Astronomy, The University of North Carolina at Chapel Hill, Chapel
Hill, NC 27599}


\vspace{-12pt}

\begin{references}
\vspace{-48pt}

\bibitem{Krav} S. V. Kravchenko, D. Simonian, M. P. Sarachik, W. Mason, 
and J. E. Furneaux, Phys. Rev. Lett.~{\bf 77}, 4938 (1996).

\bibitem{mit1} Dragana Popovi\'{c}, A. B. Fowler, and S. Washburn, Phys. Rev.
Lett. (in press), preprint cond-mat/9704249.

\bibitem{APL} K. Ismail, M. Arafa, Frank Stern, J. O. Chu, and B. S. Meyerson,
Appl. Phys. Lett.~{\bf 66}, 842 (1995).

\end{references}
\end{document}